\documentclass[reprint,
 amsmath,amssymb,
 aps,
]{revtex4-1}
\usepackage{graphicx}  
\usepackage{afterpage} 
\usepackage{dcolumn}   
\usepackage{bm}        
\usepackage{amssymb}   
\usepackage{amsmath}   
\usepackage{comment} 
\usepackage[caption=false]{subfig}
\usepackage{wrapfig}
\usepackage{float}
\usepackage{adjustbox}
\usepackage{multirow}
\usepackage{tabularx}
\usepackage{caption}

\newcolumntype{Y}{>{\centering\arraybackslash}X}

\usepackage{tikz}
\begin{document}
\title{Longitudinal crossover and the dynamics of uniform electron ellipsoids focused by linear chirp}

\author{X. Xiang}\email{xiangxuk@msu.edu}
\author{P.M. Duxbury}
\author{B. Zerbe}\email{zerbe@msu.edu}
\affiliation{Department of Physics and Astronomy, Michigan State University}

\date{\today}

\begin{abstract}
High resolution single-shot non-relativistic ultrafast electron microscopy(UEM) relies on adaptive optics to compress high intensity bunches using Radio Frequency (RF) cavities.  We present a comprehensive discussion of the analytic approaches available to characterize bunch dynamics as an electron bunch goes through a longitudinal focal point after an RF cavity where space charge effects can be large.  Methods drawn from the Coulomb explosion literature, the accelerator physics literature and the analytic Gaussian model developed for UEM are compared, utilized and extended in some cases.  In particular the longitudinal focus may occur in two different regimes, a bounce-back regime and a crossover regime; and we characterize the critical point separating these regimes in the zero-emittance model.    Results from N-particle simulations using efficient multipole methods are compared to the theoretical models revealing features requiring extensions of the analytic approaches; and in particular mechanisms for emittance growth and transfer are discussed. 

\end{abstract}

\maketitle

\section{Introduction}

Modern ultrafast microscopy has the goal of resolving sub-picosecond time periods at
sub-nanometer length scales \cite{Hall:2014_report,BES_report:2016_electron_sources}.
Consistently obtaining such resolution would allow scientists to visualize chemistry as it happens
thus opening up a deeper understanding of mechanisms at the nano-scale that are  
important to life and modern technology\cite{Miller:2014_review,Miller:2014_science_review}.  
While a number of techniques are being explored to realize such microscopy \cite{Srinivasan:2003_UED,Dwyer:2006_FED,Zewail:2006_4d_review,Ruan:2009_nanocrystallography,Sciaini:2011_review,Ischenko:2014_review2,Miller:2014_review,Miller:2014_science_review,Xu:2016_review,King:2005_review,Zewail:2010_4d,Plemmons:2015_review,Feist:2017_UEM}, weakly-relativistic ultrafast electron microscopy (UEM),
where the electron bunch has energies that are at most a significant fraction of the rest energy of the electron, 
has a number of attractive advantages.  The first advantage is the engineering fact that the device needed for such experiments
can be built on top of existing electron microscopes keeping additional engineering and expenses to a minimum. The second 
advantage is the physical fact that
the use of strongly interacting electrons means that the number of electrons required to form an image is a relatively small number
as compared to x-rays for example \cite{King:2005_review}.  The ultimate goal is to reach the single-shot limit where the number of electrons in a bunch is large enough to form an image, but weakly-relativistic UEM also
introduces technological hurdles as the space-charge effects of a high-density probing electron bunch
is considerable at a number of points within the column\cite{Tao:2011_quantitative,Tao:2012_space_charge,Portman:2013_computational_characterization,Portman:2014_image_charge,Portman:2015_multiscale,Williams:2017_longitudinal_emittance,Zerbe:2018_coulomb_dynamics}.
These effects need to be characterized to provide an accurate model for design of high-intensity beamlines.   
In this paper we describe strong space-charge effects at a longitudinal focal point.  

While such so-called space-charge dominated regimes have been well described
by accelerator physicists for cylindrical beams\cite{Reiser:1994_book}, the weakly-relativistic
bunched nature, which can be thought of as ellipsoids with finite longitudinal extent, of the electrons in UEM requires additional tools.  Some work has already characterized such dynamics in the non-relativistic regime
near the electron source.  Models of the longitudinal evolution of the bunch 
have been developed to describe the early dynamics of a bunch within an acceleration field before the center of mass motion becomes relativistic\cite{Luiten:2004_uniform_ellipsoidal,Siwick:2002_mean_field,Qian:2002_fluid_flow,Reed:2006_short_pulse_theory,Collin:2005_broadening}; although we recently showed that the transverse dynamics should not be ignored when attempting to capture
the important aspects of the electron bunch evolution\cite{Zerbe:2018_coulomb_dynamics}.  
Furthermore, once the bunch has expanded sufficiently, it is
often argued that the internal space-charge effects become negligible; however, for weakly-relativistic UEM
the bunch is recompressed resulting in the space-charge effects becoming significant at and near
focal points where the density of the bunch is again high.  

Fortunately, there have been tools developed in the astrophysics and Coulomb explosion literature where the mean-field effects of
a uniform ellipsoidal electron bunch can be modeled through ordinary differential equations.  Specifically, Lin et al. developed
a model of gravitational collapse of an oblate ellipse that could be written as a system of differential equations for the ellipses' 
widths\cite{Lin:1965_gravitational_uniform_ellipsoid_collapse}.  
Similar techniques using the repulsive electrostatic force were developed by Grech et al. to model the inverse problem
of Coulomb explosion\cite{Grech:2011_coulomb_explosion}.  
Both techniques require a tractable force, and to simplify the analysis, both techniques
assumed a uniform ellipsoid throughout the bunch evolution.

Separate from these efforts, Michalik and Sipe introduced an Analytic Gaussian (AG) model that predicts not only the 
spatial width evolution but the full phase space evolution
\cite{Michalik:2006_analytic_gaussian,Michalik:2009_AG_comparison,Berger:2010_AG_focus}.
To make such a system analytically tractable, the rms emittance, which we will clearly define later, is introduced and assumed to be conserved.   The AG model is presented in the reference frame of the bunch, so it is only applicable as long as the bunch
remains non-relativistic within the lab frame.  

We argue here that the AG model is equivalent
to the much older KV envelope equations initially developed to describe the evolution of uniform ellipsoidal distributions\cite{Kapchinskij:1959_KV}.
Sacherer provided a simple perspective that showed that the KV envelope equations could be derived from
basic, fundamental statistical considerations with applications of the mean-field force present from 
a uniform distribution\cite{Sacherer:1971_envelope}, and the mathematical form of the 
AG model may be derived from similar considerations again assuming emittance
conservation.  We provide such a derivation later in this manuscript.

In this work, we have three primary goals: 1.) We extend the model of Grech et al. to capture focussing events,
2.) we place the envelope equations
within the context of the UEM literature, and 
3.) we determine to what extent and to what effect non mean-field interaction effects lead to violation of the 
emittance conservation assumption.  
We start by extending the model employed by Grech et al. to include linear initial momentum-spatial correlations, aka chirp, 
allowing us to extend this standard approach to focused charged bunches in Sec.\ref{sec_MFT}. 
We call this the modified Coulomb explosion (MCE) model in the text.
We find that this MCE model 
naturally leads to the concept of a critical chirp that describes a collective behavior transition for particles 
within this model.  Next, in Sec.\ref{sec_SSA}, we derive the AG formalism 
from a statistical vantage point assuming a linear force.  We explicitly demonstrate 
how the Gaussian assumption differs from the uniform 
assumption only by a constant that can be absorbed into the number of particles in the bunch if the model is used to represent experimental data for example (see Appendix A). 
Further we point out that the envelope equations we derive from this statistical perspective are a generalization of the 
MCE model.  This observation allows us to partially disentangle 
the effects of the self force and emittance 
on the predicted dynamics of the bunch, and we analyze some important physics of bunch evolution 
using this insight. 
Finally, in Sec.\ref{sec_MD}, the theoretical predictions are compared with 
$N$-particle simulations, which are able to model scattering events that should alter the bunch emittance during the simulation. 
Consistent with previous theory\cite{Reiser:1994_book}, we show that emittance is transferred from hotter to colder dimensions; 
however, we also show that emittance increases almost simultaneously in both the transverse and longitudinal directions 
around crossover when the initial chirp is larger than the critical chirp.  We note that this can not be explained through 
the standard mechanism of heat transfer; and we postulate two mechanisms that provide insight into the
dynamics of emittance growth near crossover.

\section{Spatial evolution\label{sec_MFT}}
We revisit Grech et al.'s model for Coulomb explosion.  Broadly, this model assumes that the force acting
on a particle within the uniform ellipsoidal ensemble is the mean-field force calculated by the application of 
Laplace equations to a uniform distribution of electrons.  The modification we introduce is an initial linear relationship between
the initial position and the initial velocity of the particle, which we will call the ``chirp", and this modification naturally 
leads to the identification of a critical chirp that demarcates two qualitatively different regimes of bunch behavior within this model.

\subsection{The mean-field framework}
We first recall the well-known quadratic form of the electrostatic potential for position $(x,y,z)$ inside a uniform electron ellipsoidal bunch with semi-axes of $(a,b,c)$ and charge number density $n$ that can be obtained using Laplace's equations:
\begin{widetext}
	\begin{equation}
	V(x,y,z)=\frac{n \cdot abc \cdot e}{4 \varepsilon_0}
	\cdot \int^{\infty}_0\left( 1-\frac{x^2}{a^2+s}-\frac{y^2}{b^2+s}-\frac{z^2}{c^2+s} \right) \frac{ds}{\sqrt{(a^2+s)(b^2+s)(c^2+s)}},\label{eq:pot}
	\end{equation}
\end{widetext}
where $\varepsilon_0$ is the vacuum permittivity. We assume rotational symmetry about the $z$ axis enabling us to 
introduce the radial coordinate $r = \sqrt{x^2+y^2}$. Although the detailed calculations 
below are performed specifically for prolate ellipsoids ($a=b<c$), similar results are valid for general uniformly charged ellipsoidal bunches. 

The electrostatic field may be obtained from Eq. (\ref{eq:pot}) using $\vec{E} = -\vec{\nabla} V$.  Due to the symmetry,
the angular portion of the field is $0$.  Thus the electric field may be written as:
\begin{equation}
\vec{E}(r,z) = E_r(r) \hat{r} + E_z(z) \hat{z}
\end{equation}
with $\hat{r}$ and $\hat{z}$ representing the radial and longitudinal unit vectors, respectively, and
\begin{subequations}\label{eqn_Efield}
	\begin{align}
	E_r(r) &= \frac{n e}{2 \varepsilon_0} \xi_r(\alpha) \cdot r \\
	E_z(z) &= \frac{n e}{2 \varepsilon_0} \xi_z(\alpha) \cdot z
	\end{align}
\end{subequations}
where $\alpha = a/c$ is the ellipsoid aspect ratio and the corresponding geometry coefficients $\xi_r(\alpha)$ and $\xi_z(\alpha)$ are
\begin{subequations}\label{eqn_xi}
	\begin{align}
	\xi_r(\alpha) &= \alpha^2 \int^{\infty}_0 \frac{ds}{(\alpha^2+s)^2(1+s)^{1/2}} \\
	\xi_z(\alpha) &= \alpha^2\int^{\infty}_0 \frac{ds}{(\alpha^2+s)(1+s)^{3/2}}
	\end{align}
\end{subequations}

The linear relation between the electric field felt by a particle and the particle's position results in the 
preservation of the uniformity of the ellipsoidal bunch. This greatly simplifies our analysis as the formulation 
presented above applies to the bunch for all time and the evolution reduces to the determination of two degrees of  
freedom. Specifically, the temporal evolution of the entire bunch can be represented by the evolution of two unit-less 
scaling functions, $R(t)$ and $Z(t)$, i.e. the trajectory of any particle with initial position $(r_0,z_0)$ inside the uniform ellipsoid is given by $(r_0 R(t), z_0 Z(t))$, where $R$ and $Z$ are independent of the initial position $(r_0,z_0)$. Thus, the parameters for describing the bunch changes accordingly: (i) the semi-axis of the ellipsoids can be written as $(a, c) = (a_0 R, c_0 Z)$, (ii) the transient aspect ratio can be written as $\alpha(t) = \alpha_0 \cdot R/Z$, (iii) the number density can be derived using conservation of charge $N_{total} = n_0 \cdot (4\pi/3)a_0^2 c_0 = n(t) \cdot (4\pi/3) a^2 c$ giving $n(t) = n_0/(R^2 Z)$, and (iv) the spatial variance of the bunch change to $\sigma^2_z(t) = \sigma^2_{z0} \cdot Z^2$ and $\sigma^2_r(t) = \sigma^2_{r0} \cdot R^2$.  Therefore, it should be apparent
that any parameter in the problem can be determined from $R$ and $Z$, which we set out to determine for all time. 

In the non-relativistic limit, the equations of motion (EOM) of a particle inside the field determined in Eq. (\ref{eqn_Efield})
can be simply determined using $\ddot{\vec{x}} = \frac{q}{m} \vec{E}$.  These EOM reduce to two dimensionless ordinary 
differential equations (ODEs) for our scaling parameters:
\begin{subequations}\label{eqn_EOM_RZ}
	\begin{align}
	\frac{d^2 R}{d\tau^2} &= \frac{\xi_r(\alpha)}{R Z} \\
	\frac{d^2 Z}{d\tau^2} &= \frac{\xi_z(\alpha)}{R^2}\label{eq:d2Zdt2}
	\end{align}
\end{subequations}
with unit-less reduced time, 
\begin{equation}
\tau = t\cdot \sqrt{\frac{e^2 n_0}{2\varepsilon_0 m}} = t \cdot \Omega_0
\end{equation}
and electron mass $m$. Notice that: (i) the time scaling factor $\Omega_0 = \frac{1}{\sqrt{2}} \omega_{p0}$ 
where $\omega_{p0}(n_0)=\sqrt{\frac{e^2 n_0}{\varepsilon_0 m}}$ is the initial plasma frequency and (ii) the geometry 
coefficients $\xi_r$ and $\xi_z$ solely depend on the aspect ratio $\alpha$ rather than specific value of $a$ and $c$. This means 
that starting with the same initial conditions for the ODEs, bunches with the same initial aspect ratio $\alpha_0$ but different 
initial density $n_0$ will lead to identical behaviors only differing by the time scaling factor $\Omega_0$ determined by the 
initial number density $n_0$.  
Eq. (\ref{eqn_EOM_RZ}) are more or less the ODE's used by Lin et al.\cite{Lin:1965_gravitational_uniform_ellipsoid_collapse} 
and Grech et al.\cite{Grech:2011_coulomb_explosion} except we have 
scaled the time to be more general, so the model we have presented so far
does not significantly differ from those works.

\subsection{Initial conditions}
The behavior predicted by a specified system of non-chaotic ODE's is entirely determined by its initial conditions,
and the initial conditions we consider are
\begin{subequations}\label{eqn_cold_init}
	\begin{align}
	R(\tau=0) &= 1 \label{eqn_cold_init_R} \\
	Z(\tau=0) &= 1 \label{eqn_cold_init_Z} \\
	\left.\frac{dR}{d\tau}\right\vert_{\tau=0} &= -\nu_r^\ast \label{eqn_cold_init_vR}\\
	\left.\frac{dZ}{d\tau}\right\vert_{\tau=0} &= -\nu_z^\ast \label{eqn_cold_init_vZ}
	\end{align}
\end{subequations}
where $\nu_i^{\ast}$ is trivially proportional to the linear chirp.  We call $\nu_{z}^\ast$ the reduced 
longitudinal chirp, and
its proportionality to the linear chirp can be obtained by noting
$p_z(z_0)= m \cdot (\nu^{\ast}_z\Omega_0 \cdot z_0) = m \mathcal{C}_z \cdot z_0$ 
where $\mathcal{C}_z$ is the longitudinal linear chirp.
Notice that Eqs. (\ref{eqn_cold_init_R}) and (\ref{eqn_cold_init_Z}) represent the initial scaling of the ellipsoid and are by definition set to $1$ as these parameters represent the scaling of the transverse and longitudinal dimensions, respectively, from their 
initial values. On the other hand, Eqs. (\ref{eqn_cold_init_vR}) and (\ref{eqn_cold_init_vZ}) represent 
the initial rate of change of the scale functions $R$ and $Z$, which can be roughly thought of as the velocity of the expansion.
Lin et al. and Grech et al. set 
$\nu_r^\ast = \nu_z^\ast = 0$ to model gravitation collapse and Coulomb explosion, respectively, 
where the bunch is assumed to start from rest. 
The Coulomb explosion results were 
found to be in good agreement with molecular dynamics (MD) simulations for time-dependent energy 
distributions and particle-in-cell (PIC) simulations for temporal ellipsoid radii evolution\cite{Grech:2011_coulomb_explosion}.
For our purposes, we assume $\nu_r^\ast = 0$ and $\nu_z^\ast \ne 0$ to model the effect of a longitudinal lens,
e.g. a RF cavity.  Specifically, notice that if the reduced longitudinal 
chirp is positive, i.e. $\left.\frac{dZ}{d\tau}\right\vert_{\tau=0}$ is negative, 
$Z$ will initially decrease, and the bunch will be focused in the longitudinal direction.
In summary, the focusing process of a uniform charged ellipsoid is entirely determined by its initial aspect ratio and its 
reduced longitudinal chirp as the density of initial bunch determines only the time scale of the evolution.  We call this general form of Grech et al.'s model the modified Coulomb explosion (MCE) model.
 
In particular note that the reduced longitudinal chirp is unitless while the longitudinal chirp has units of inverse time.
This is because the reduced chirp 
is the actual chirp scaled by $\Omega_0$, and this cancels the time units.  As $\Omega_0$ depends solely on density, 
the reduced chirp is more general as the density determines the time scale and therefore the ODE represents the
interplay between the geometry and the electrostatic force.  However, if the density is not important in our discussion of 
some physical observation, 
we will often drop the ``reduced" when discussing the chirp as the statement should apply to both the reduced chirp as well
as the actual chirp. 

\subsection{Critical reduced chirp}

\begin{figure}
	\includegraphics[width=0.85\linewidth]{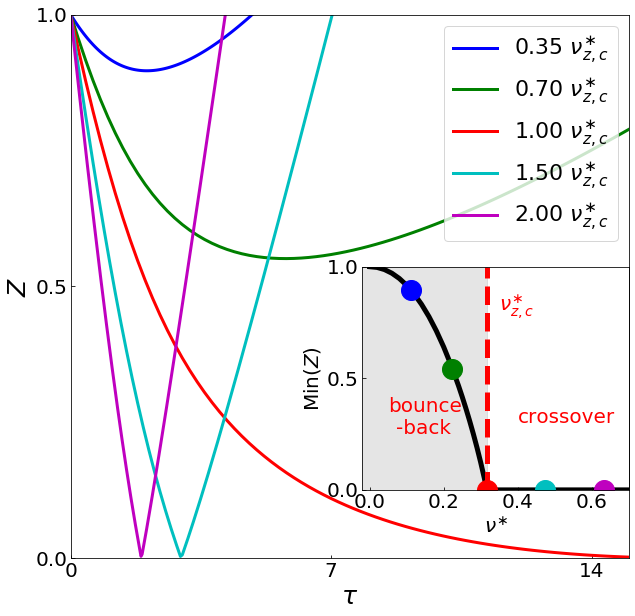}	
	\caption{\label{fig_above_and_below_CC} Longitudinal width evolution $Z = Z(\tau)$ of prolate ellipsoids with $(\alpha_0=10/75)$ driven by different initial chirps in numeric solutions of the MCE model ranging from below the critical chirp ($0.35 ~\nu_{z,c}^{\ast}$) to well above ($2.0 ~\nu_{z,c}^{\ast}$). The sub-graph shows the dependence of minimum width on initial reduced chirp. 
	The red dot represents the critical value $\nu_{z,c}^{\ast}$ for the particular $\alpha_0 = {10}/{75}$, 
	and the bounce-back and crossover regimes a separated by the vertical, dashed, red line at this point.}
\end{figure}

As the effect of aspect ratio on the evolution has been well studied previously\cite{Grech:2011_coulomb_explosion}, we examine the effect of the reduced longitudinal chirp on the bunch focusing of a pre-specified aspect ratio, $\alpha_0=10/75$.  Specifically, we are interested
in modeling the bunch reaching a minimum in longitudinal extent after $\tau = 0$ which
occurs when $-\nu_z^\ast < 0$.  We define the time to focus, 
$\tau_f$, as the unitless time at which the bunch reaches its minimum longitudinal width. 
As can be seen in Fig. \ref{fig_above_and_below_CC},
$\tau_f$ is a function of the reduced chirp, i.e. $\tau_f = \tau_f(\nu_z^\ast)$.  

Furthermore, define $Z_f$ to be the longitudinal scaling parameter at the focal point,
i.e. $Z_f = Z(\tau_f) = min(Z(t))$.  Moreover notice that $Z(t) \ge0$, so $Z_f \ge 0$; in fact, for sufficiently large  
reduced chirps $Z_f = 0$ as can be seen in Fig. \ref{fig_above_and_below_CC}.  This is because the evolution of the 
longitudinal scaling parameter, seen in Eq. (\ref{eq:d2Zdt2}), is dependent only on $1/R^2$, and $R > 1 $ in our model.  This
means that if $\nu_z^\ast$ is sufficiently large, the initial longitudinal chirp overcomes the repulsion of the
electrostatic force and the bunch briefly collapses to a two-dimensional object at the focal point. 
We call the smallest magnitude of the reduced chirp for satisfying this condition the critical chirp, $\nu^{\ast}_{z,c}$. 
Notice that $\tau_f(\nu_{z,c}^\ast) = \infty$ and that $\frac{d \tau_f}{d \nu_{z}^\ast} \begin{cases}>0,& \nu_{z}^\ast < \nu_{z,c}^\ast\\ <0,& \nu_{z}^\ast > \nu_{z,c}^\ast\end{cases}$ again as can be seen in 
Fig. \ref{fig_above_and_below_CC}.

In other words, the behavior of the model can be partitioned into two categories characterized by whether the initial longitudinal chirp
is greater than or less than the critical chirp.  More specifically, as the magnitude of focusing chirp is increased from $0$, 
the minimum width of the bunch decreases and the time to focus increases. This trend continues until the critical 
chirp is reached where the corresponding time to the focal point becomes infinitely large, i.e. $\tau_f \rightarrow \infty$
as $\nu_z^\ast \rightarrow \nu^{\ast}_{z,c}$. 
Above the critical chirp, the bunch will overcome the Coulomb repulsion and be compressed through a longitudinal crossover as electrons starting from one side of the bunch cross the center of mass and then begin to expand on the other side. We refer to this as the ``crossover" regime, and in this regime further increasing the chirp has no effect on the $0$ minimum width but decreases the time to focus. In contrast, we call the regime below the critical chirp the ``bounce-back" regime as a particle within the bunch
with such a chirp follows a trajectory that reverse its initial direction.

The crossover event adds complexity to simulations of the model.  Specifically, 
the linearity of both the force and the velocities of the particles in the model indicates that all the crossover 
incidents happen simultaneously across the bunch at $\tau_f$ within the crossover regime, 
creating a 2D singularity in the EOM with $Z \rightarrow 0$.  Before the crossover, the chirp is negative while after the crossover
the chirp becomes positive.  As the force in the $z$-direction is very small due to geometric considerations, the speed of the particles do not change substantially, just the sign of the linear relationship in phase space.  
This necessitates careful treatment of the chirp through the crossover event.  We accomplish this treatment 
by using a small time step to propagate the EOM up until $Z$ goes below zero.  As $Z$
is a scale, the negative sign has no physical meaning and indicates that crossover occurred within the previous time step.  
So, we stop the simulation and flip the value of both longitudinal position scaling, $Z$, and longitudinal momentum, $p_z$.  After this, the same EOM are used to integrate the parameters.  
In effect, this skips the singularity by an infinitesimal step size in time. In addition, this also implies that the crossover case, where $Z$ will pass through $0$ in this fashion,
will have a sudden change in longitudinal chirp as compared to the bounce-back case where such $Z$ does not pass through $0$ and
the chirp instead smoothly changes due to the effect of the repulsive mean-field force. 

Analogous to our longitudinal treatment, a radial chirp can also be added by setting $\nu_r^\ast$ 
in Eq.\ref{eqn_cold_init_vR} to a non-zero value.  
Furthermore, this treatment may be combined with the longitudinal chirp to model bunches focused in both degrees of freedom simultaneously -- a treatment that is outside of the scope of this paper. 
However, in contrast to the longitudinal dimension, this model predicts that there is no such critical chirp or crossover in radial focusing. This occurs because of $d^2R/d\tau \propto 1/R$ as can be seen in Eq. (\ref{eqn_EOM_RZ}).  
This indicates that the force in the transverse direction diverges as the bunch focuses radially preventing the singularity in 
the longitudinal scaling parameters 
seen when only the longitudinal direction is focused. 

In the general situation, this difference between being able to 
focus through a singularity longitudinally but not transversely is a result of attempting to focus two dimensions, 
i.e. $\hat{x}$ and $\hat{y}$, simultaneously. Focusing in more than one dimension in this model is not 
possible even when all dimensions are treated separately as the Coulomb repulsion on one 
dimension is inversely dependent on the widths of the other two dimensions. In other words, there is only the bounce-back 
regime when more than one dimension is focussed concurrently.
We will later (in Sec.\ref{sec_SSA}) discuss how emittance influences the minimum width of the bunch,
and this statistical measure reintroduces the ability of particles to crossover even when the bunch is in the bounce-back 
regime.  For the rest of this manuscript, though, we will only focus on the longitudinal focusing where both 
the crossover and bounce-back regimes are accessible in the model.

\begin{figure}[h]
	\includegraphics[width=0.8\linewidth]{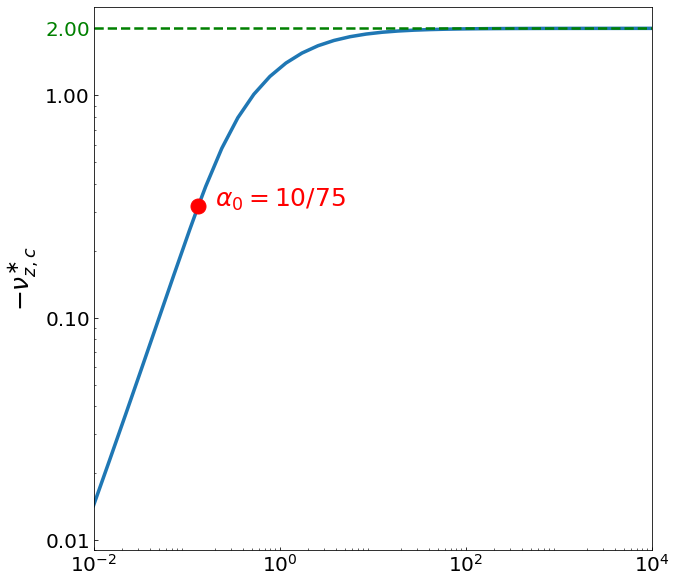}
	\caption{\label{fig_CC_alpha} Dependence of critical reduced longitudinal 
	chirp, $\nu_{z,c}^{\ast}$, on the initial aspect ratio, $\alpha_0$. The green dashed line represents the 
	horizontal asymptote, $\eta_{z,c}^{\ast}(\alpha_0 \rightarrow \infty) = 2$. 
	The red dot represents the aspect ratio of our MD simulation.}
\end{figure}

One important feature of the critical reduced chirp, $\nu_{z,c}^{\ast}$, is its exclusive dependence on the initial aspect ratio $\alpha_0$. This fact stems from the governing EOM solely depending on the aspect ratio. In Fig. \ref{fig_CC_alpha}, we present the reduced critical chirp as a function of the initial aspect ratio.  Specifically, note that for large $\alpha_0$ often referred to as the 
``pancake" regime \cite{Luiten:2004_uniform_ellipsoidal,Siwick:2002_mean_field,Qian:2002_fluid_flow,Reed:2006_short_pulse_theory,Collin:2005_broadening,Zerbe:2018_coulomb_dynamics}, 
we always have $\alpha \gg 1$, where the geometry coefficients $\xi_r$ and $\xi_z$ can be approximated in closed forms, with 
\begin{subequations}
\begin{align}
\xi_r(\alpha \rightarrow \infty) &\simeq \pi/(2\alpha) \rightarrow 0 \\
\xi_z(\alpha \rightarrow \infty) &\simeq 2-\pi\alpha^2(\alpha^2-1)^{-3/2} \rightarrow 2
\end{align}
\end{subequations}
Therefore, the longitudinal motion can be treated as the elementary constant acceleration kinematic equation:
\begin{equation}
Z(\tau) = Z_0 - \nu_{z,c}^{\ast} \tau + \frac{1}{2} \cdot \xi_z(\infty) \cdot \tau^2
\end{equation}
with $\tau_f = \nu_{z,c}^{\ast}/\xi_z(\infty)$ and $Z(\tau_f)=0$. In such cases, the critical reduce chirp $\nu_{z,c}^{\ast}(\alpha_0 \rightarrow \infty)$ reduces to $2$, as shown in Fig. \ref{fig_CC_alpha}.

\section{Envelope equations\label{sec_SSA}}
In this section, we present a brief derivation of the envelope equations, and we 
compare this model to the Analytic Gaussian (AG) formalism.  
Our derivation is essentially identical to Sacherer's derivation of the Kapchinsky-Vladimirsky envelope equations\cite{Sacherer:1971_envelope} and closely follows our recent derivation presented with 
our discussion of the sample perspective\cite{Zerbe:2020_1d_emittance}. 

\subsection{Derivation}
We first introduce the statistics of the bunch and their dynamics. In each degree of freedom ($x$, $y$, $z$), 
we need three quantities to describe the second order statistics of the phase space: 
$s_i$, $s_{p_i}$ and $s_{i,p_i}$, with $i=T,z$ for the transverse ($T = x, y$) or longitudinal direction. 
The basic statistics are then
\begin{subequations}
	\begin{align}
		s^2_i &= \overline{i^2} - \bar{i}^2 \\
		s^2_{p_i} &= \overline{p_i^2} - \bar{p}_i^2  \\
		s_{i,p_i} &= \overline{i p_i} - \bar{i} \bar{p}_i 
	\end{align}
\end{subequations}
where the bar operator indicates the mean, e.g. $\overline{x p_x} = \frac{1}{N} \sum_{j=1}^N x_j p_{x,j}$.
As the number of particles is a constant, derivatives commute with sums, and derivatives of products can be
determined by the chain rule,
we have $\frac{d}{dt}\bar{a} = \overline{\frac{da}{dt}}$ and it is straightforward to show
\begin{align}
  \frac{d}{dt} s_{a,b} &= s_{\frac{da}{dt},b} + s_{a,\frac{db}{dt}}
\end{align}
Thus the time derivatives of our phase-space statistics are
\begin{subequations}
	\begin{align}
		\frac{ds^2_i}{dt} &= \frac{2}{m}s_{i,p_i} \\
		\frac{ds_{i,pi}}{dt} &= s_{i,F_i} + \frac{1}{m}s^2_{p_i}\\
		\frac{ds^2_{pi}}{dt} &= 2s_{p_i,F_i}
	\end{align}
\end{subequations}
assuming non-relativistic dynamics.

This system of equations is equivalent to the system presented in the AG model.  
Specifically, introduce the AG parameters 
\begin{subequations}
	\begin{align}
		\sigma^{AG}_i &= s^2_i \\
		\gamma^{AG}_i &= s_{i,p_i} \\
		\eta^{AG}_i &= s^2_{p_i} - \frac{s_{i,p_i}^2}{s^2_i} 
	\end{align}
\end{subequations}
Subbing these parameters into our ODE we obtain
\begin{subequations}\label{eqn_SSA_dynamics}
	\begin{align}
	\frac{d \sigma_i^{AG}}{dt} &= \frac{2}{m} \gamma_i^{AG}  \\
	\frac{d \gamma_i^{AG}}{dt} &= \frac{1}{m}\left( \frac{{\gamma_i^{AG}}^2}{\sigma_i^{AG}} + \eta_i^{AG} \right) +  s_{i,F_i} \label{eq:gamma} \\
	\frac{d \eta_i^{AG}}{dt} &= -\frac{2 \gamma_i^{AG} \eta_i^{AG}}{m \sigma^{AG}_i} + \frac{2}{\sigma^{AG}_i} \left( \sigma^{AG}_i s_{p_i,F_i} - \gamma_i^{AG} s_{i,F_i} \right)\label{eq:eta}
	\end{align}
\end{subequations}
This system of equations differs from Michalik and Sipe's published system of ODE's in two ways: 1.)  
in Eq. (\ref{eq:gamma}) we have 
$s_{i,F_i}$ instead of Michalik and Sipe's $\frac{1}{4\pi\epsilon_0} \frac{N e^2}{6 \sqrt{\pi} \sigma_i} L_i\left(\frac{s_z}{s_T}\right)$ 
where $L_i\left(\frac{s_z}{s_T}\right)$ is an integral we examine in detail in the Appendix A
 and 2.) we include 
$\frac{2}{\sigma^{AG}_i} \left( \sigma^{AG}_i s_{p_i,F_i} - \gamma_i^{AG} \sigma_{i,F_i} \right) = \frac{1}{s_i^2} \frac{dm^2 c^2\epsilon_{i,p_i}^2}{dt}$ where  $\epsilon_{i,p_i}^2 = \frac{1}{m^2c^2} (s_i^2 p_i^2 - s_{i,p_i}^2) = \sigma_i^{AG} \eta_i^{AG}$
in Eq. (\ref{eq:eta})
which Michalik and Sipe treat as $0$.
As noted by our notation, the latter additional term represents the effect of the change of emittance on the bunch evolution.  
We believe that that Michalik and Sipe's
assumption of self-similar evolution leads to this term vanishing and hence emittance being conserved.
This assumption is not strictly true as
we have recently shown that the Gaussian
distribution does not evolve self-similarly under the Coulomb force\cite{Zerbe:2018_coulomb_dynamics}; 
nonetheless, self-similarity may be a reasonable assumption in order to approximate spatial statistics.  

A single assumption reproduces the published form of the Analytic Gaussian model from Eq. 
(\ref{eqn_SSA_dynamics}); which is to assume that the force on a particle in the AG model
is linear and can be written as  
\begin{align}
  F_i(i) &= \frac{1}{4\pi\epsilon_0} \frac{N e^2}{6 \sqrt{\pi} (\sigma_i^{AG})^{3/2}} L_i(\xi) i
\end{align}
This assumption leads to 
$s_{i,F_i} = \frac{1}{4\pi\epsilon_0} \frac{N e^2}{6 \sqrt{\pi} \sqrt{\sigma_i^{AG}}} L_i(\xi)$ 
and 
$\frac{2}{\sigma^{AG}_i} \left( \sigma^{AG}_i s_{p_i,F_i} - \gamma_i^{AG} s_{i,F_i} \right) = 0$ 
reproducing Michalik and Sipe's published system of ODE's.  

However, this linear force assumption has 2 somewhat subtle, and related, problems.  
The first has to do with the description of the relationship between the
position of a particle in the distribution and the force it experiences
on average.  The line of best fit has slope $\frac{s_{i,F_i}}{s_i^2}$, so $s_{i,F_i}$ can be thought of as the
slope of the best fit line times the spatial variance.  While the slope of the line for a Gaussian is essentially
described by the force in Michalik and Sipe's AG model, 
we have already mentioned that 
the assumption that the distribution will remain Gaussian has been found to be incorrect\cite{Zerbe:2018_coulomb_dynamics}.  Specifically,  the slope of the best fit line is specific to the given distribution, and as the Gaussian distribution evolves toward
a ringed distribution, the slope of this line changes partially just in response to this change in distribution.
This issue can be partially avoided theoretically by assuming a uniform distribution that does continue to be uniform as it evolves 
--- at least in the continuum, mean-field, non-relativistic, zero-emittance
limit.  For this work, we assume $F_i(i) = (m \omega^2_{p_0}/2) \xi_i(\alpha) i$ which leads to the envelope equations
we use as well as the equations used by Sacherer\cite{Sacherer:1971_envelope}.  
The difference between the AG model and these uniform envelope equations can be mathematically shown to be nothing more than
a difference in a unitless constant  relating the force used in the analysis of Michalik and Sipe, the force we use. In the Appendix, we calculate this constant and find that it is only $1.05$ indicating that these models are essentially equivalent.
Because these two models differ by only about 5\% in the force, 
either set of equations can be used in most applications to experiment.

However, the linear assumption results in a more serious issue.  Specfically the force during an $N$-particle simulation
differs from the linear approximation.  This is even true for the uniform distribution although the more consistent
non-linearities of the Gaussian distribution result in more significant deviations. 
So  while the $s_{x,F_i}$ can be captured in many situations with the mean-field force, 
it is important to understand these deviations especially for the
term relating the momentum and the force, 
$\frac{m^2 c^2}{s_i^2} \frac{d\varepsilon_{i,p_i}^2}{dt} = \frac{2}{s^2_i} \left( s^{2}_i s_{p_i,F_i} - s_{i,p_i} s_{i,F_i} \right)$,
cannot be likewise captured.  
As the Gaussian distribution results in significant deviations between the linear force and the mean-field force, the
rms emittance has additional effects on the evolution of the emittance growth than the uniform distribution's purely stochastic
driven emittance changes.    
This second point has not been examined in the literature, and we begin to evaluate
the stochastic driven aspects of emittance growth in this manuscript by investigating the emittance change during simulation
as compared to the uniform envelope equation predictions.  

In terms of understanding the physics, it is convenient to introduce a variable representing the average
local variance in the momentum,
\begin{align}
  \eta_i^2 &= s_{p_i}^2 - \frac{s_{i,p_i}^2}{s_i^2}
\end{align}
Notice that $\eta_i^2$ in our notation is equivalent to $\eta_i^{AG}$ but that our $\eta_i$ has units of momentum
and $\frac{1}{2m} \eta_i^2$ has units of energy.
Furthermore, the emittance can be written as $\epsilon_{i,p_i} = \eta_i s_i$.
With this notation, our system of ODEs becomes
\begin{subequations}\label{eqn_SSA_dynamics}
	\begin{align}
	\frac{d s^2_i}{dt} &= \frac{2}{m} s_{i,p_i}  \\
	\frac{d s_{i,p_i}}{dt} &= \frac{1}{m}\left( \frac{s^2_{i,p_i}}{s^2_i} + \eta_i^2 \right) +  \mathcal{K}_{Fi}(\alpha) s^2_{i} \label{eqn_gamma} \\
	\frac{d \eta_i^2}{dt} &= -\frac{2 s_{i,p_i} \eta_i^2}{m s^2_i} \label{eqn_SSA_eta}
	\end{align}
\end{subequations}
where we have used the force expression, $F_i = \mathcal{K}_{Fi}(\alpha)\cdot i = (m \omega^2_{p_0}/2) \xi_i(\alpha) i$.

Although the above derivations have been performed for the Coulomb interaction, we would like to stress that the same conclusions can be drawn for any interaction that leads to linear dependence between force and position. 
Additionally, the generalization to any general ellipsoid is simple using three degrees of freedom with $i=X,Y,Z$ and corresponding geometry coefficients $(\xi_x,\xi_y,\xi_z)$ as functions of the ratio between three axes $(s_x:s_y:s_z)$ as we recently
pointed out\cite{Zerbe:2020_1d_emittance}. 

\subsection{Dynamics of the envelope equations}

\begin{figure}[h]
	\includegraphics[width=0.9\linewidth]{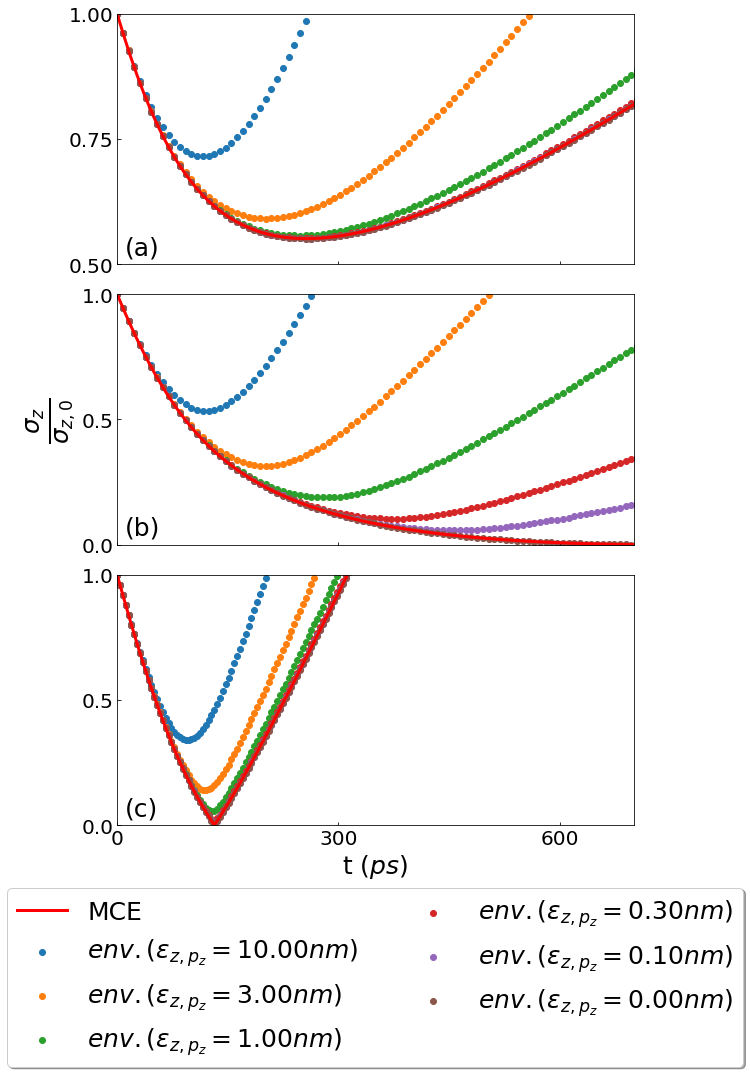}
	\caption{\label{fig_Zt_SSA} Longitudinal width evolution, $\sigma_z$, divided by the initial longitudinal width, $\sigma_{z,0}$,
	of prolate ellipsoids with ($\alpha_0=10/75$) focused by different initial 
	chirps: (a) $0.7~\nu^{\ast}_{z,c}$, (b)$1.0 ~\nu^{\ast}_{z,c}$, (c) $1.5~\nu^{\ast}_{z,c}$. 
	In each figure, the red solid line represents the prediction from the MCE model and the
	dotted lines represent the envelope equations with different emittance ranging from $0$ to $10$ nm. Notice
	that the envelope equations with zero emittance and the MCE model are in perfect agreement. Also notice that 
	(1) increasing the emittance makes the waist larger and moves it earlier and  
	(2) the evolution of the width statistic is more
	responsive to emittance when the chirp is in the vicinity of the critical chirp. }
\end{figure}

Notice that the non-interacting bunch model can be obtained by setting $\mathcal{K}_{Fi}(\alpha)  = 0$ in 
Eq. (\ref{eqn_SSA_dynamics}).  Further, notice that the MCE model gives identical predictions to the zero-emittance limit of 
the envelope equations above, as can be seen in in Fig.\ref{fig_Zt_SSA}. 

The bunch dynamics can be better understood analytically by investigating 
the dynamics of the linear chirp, $\mathcal{C}_i = \frac{s_{i,p_i}}{s_i^2}$,
\begin{equation}
	\frac{d}{dt}\mathcal{C}_i = \mathcal{K}_{Fi}(\alpha) + \frac{\varepsilon^2_i}{m  (s^2_i)^2 }\label{eq:dC_idt}
\end{equation}
That is, the chirp evolution is influenced by two ``forces": (i.) the self-interaction within the bunch modeled by 
$\mathcal{K}_{Fi}(\alpha) i $ and (ii.) the emittance of the bunch inversely 
scaled by the spatial standard variance squared, i.e. $\frac{\varepsilon^2_i}{m  (s^2_i)^2 } i$. 
This realization leads us to identify three ways to investigate the physics of the bunch dynamics: 1.) the non-interacting model where the
``forces" are determined solely by the emittances, 2.) the MCE model or equivalently the
envelope equations with $0$ emittance whose bunch dynamics are driven entirely by the internal Coulomb repulsion, and 3.)
the full envelope equations where the two effects and their interaction effect may be examined.  We discuss the effects of
these contributions in the rest of this section.

Notice in Eq. (\ref{eq:dC_idt}), the emittance term is always non-negative.  Thus, as emittance is increased, the negative chirp will
become zero faster --- that is, a larger emittance results in $t_f$, the time to the minimum spatial width, being smaller or equivalently
the waist of the focusing process appearing earlier than the prediction from the zero emittance model.  This can be seen in 
Fig. \ref{fig_Zt_SSA} where the width trajectories largely follow one another before the larger emittance predictions
break off and reach their 
minima at a short time later.  While there is a small effect of this effective emittance ``force" on the bunch width, to first order, 
it is primarily the shift in the time to focus that increases the size of the waist.

\begin{figure}[h]
	\includegraphics[width=0.85\linewidth]{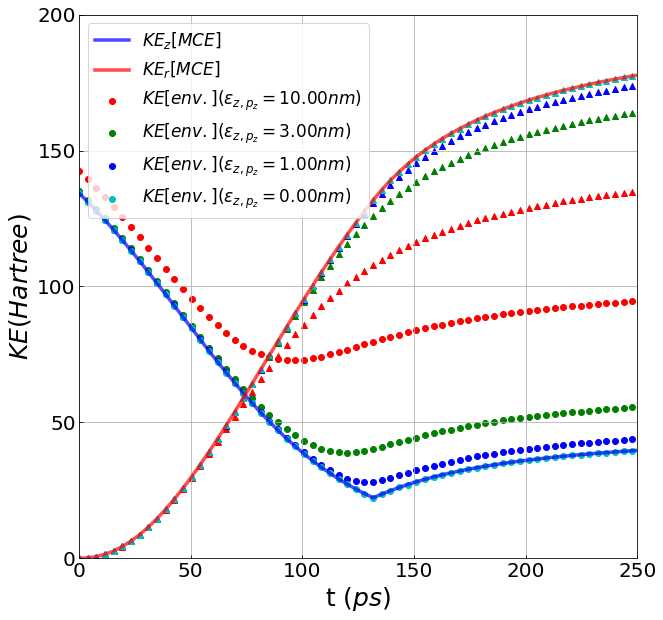}
	\caption{\label{fig_KE_SSA} The longitudinal and transverse kinetic energy for the crossover case 
	($1.5 ~ \nu^{\ast}_{z,c}$ corresponding to panel (c) in Fig.\ref{fig_Zt_SSA}), 
	with solid lines for the MCE model and dotted lines for the envelope
	equations with different emittance (circle for $KE_z$ and triangle for $KE_T$). The sudden change of direction for the slope 
	of the MCE model prediction of the longitudinal kinetic energy comes from the sign flip of chirp discussed in Sec.\ref{sec_MFT}.}
\end{figure}

The kinetic energy can also be modeled through the envelope equations.  The kinetic energy
can be exactly written as $KE = \sum_i \frac{p_i^2}{2m}$ and can be decomposed into
$KE = KE_x + KE_y + KE_z$ where $KE_i = \frac{N}{2m} (p_{i,CoM}^2 + s_{p_i}^2)$ and where $p_{i,CoM}$ is the 
momentum of a single particle at the center of mass of the bunch in the $i^{th}$ direction. 
Assuming the center of mass momentum doesn't change, 
the kinetic energy evolution along the $i^{th}$ dimension can be written as:
\begin{align}
	\frac{d}{dt} KE_i &= \frac{N}{2m} \frac{d s_{p_i}^2}{dt}\nonumber\\
	                          &=  \frac{N}{m} s_{i,p_i} \mathcal{K}_{Fi}(\alpha),
\end{align}
in any linear model.  That is, the kinetic energy can be transferred via the mean-field force into or out of the potential. 
Furthermore, as the different components ($x$, $y$, and $z$) can be independently controlled 
so that in effect energy is being transferred into the potential by one component but out of the potential by another,
this mechanism can lead to kinetic energy width transfers between the dimensions. 
The result of this mechanism can be seen in Fig. \ref{fig_KE_SSA} for a bunch in the crossover regime.
Furthermore, notice that the effect of the emittance can also be seen Fig. \ref{fig_KE_SSA}.  Specifically, the transfer of kinetic
energy between the components is reduced by increasing the emittance.  
This occurs as the non-zero emittance results in the minimum width being both larger and occurring
earlier than the MCE model.  In turn, this larger size reduces the forces experienced reducing this transfer.
Moreover, the earlier focal time results in the transfer finishing earlier, and all of these effects are factors in determining the
amount of energy transferred between dimensions.  We will provide more details of this aspect of the model in future publications.

\section{$N$-particle Simulations\label{sec_MD}}
Now that details of the models are understood, we compare the models to $N$-particle simulations. While the models 
describe the evolution of the bunch under specific conditions, i.e. conserved emittance for the envelope equations 
and zero emittance for the MCE model, 
no such assumption is present in the $N$-particle simulations. The only assumption that we make in these simulations is that 
the bunch remains non-relativistic and thus electrostatics can be used to model the inter-particle interaction.

The simulations were conducted using in-house code. This code has been validated through comparison to 
other in-house code implementing the PIC algorithm from Warp\cite{Zerbe:2018_coulomb_dynamics}. 
This code employs the non-relativistic equations 
of motion for every electron using velocity Verlet integration where we used the Fast Multipole Method (FMM) from the 
fmmlib3d library\cite{Gimbutas:2015_fmmlib3d} to calculate the field. 
As emittance increases initially due to disorder-induced heating (DIH) \cite{Gericke:2003_dih}, the bunch 
needs to equilibrate before the focusing simulation. We first place electrons inside a simulation box with periodic boundaries 
at the target density, which is $10,000$ electrons for a prolate ellipsoid with the semi-axes of $(10 \mu m, 10 \mu m, 75 \mu m)$. 
The starting position of the electrons are randomly drawn from a uniform distribution and the starting momentum is zero. 
Then the electrons are thermalized using Particle-Particle-Particle-Mesh (PPPM) methods in 
LAMMPS\cite{Plimpton:1993_LAMMPS} (http://lammps.sandia.gov) for over $10$ 
plasma oscillation periods. At the end of thermalization, we select electrons inside the desired prolate ellipsoidal region to 
construct one sample of initial conditions. To mitigate the stochastic effects in this procedure, 
we prepare $90$ such samples. This process results in 90 ellipsoids of particles with non-zero emittance that experience 
only minor additional DIH at the beginning of the focusing simulation. We call these initial conditions ``warm'' due to the 
non-zero emittance.

\subsection{\label{sec_width_discussion}Longitudinal width and kinetic energy evolution}
Simulations were performed with 3 representative initial chirps: (i.) $0.7~\nu^{\ast}_{z,c}$ in the bounce-back regime, (ii.) $1.0~\nu^{\ast}_{z,c}$ at the critical chirp, and $1.5~\nu^{\ast}_{z,c}$ within the crossover regime. 
\begin{figure}
	\begin{center}
	\includegraphics[width=0.95\linewidth]{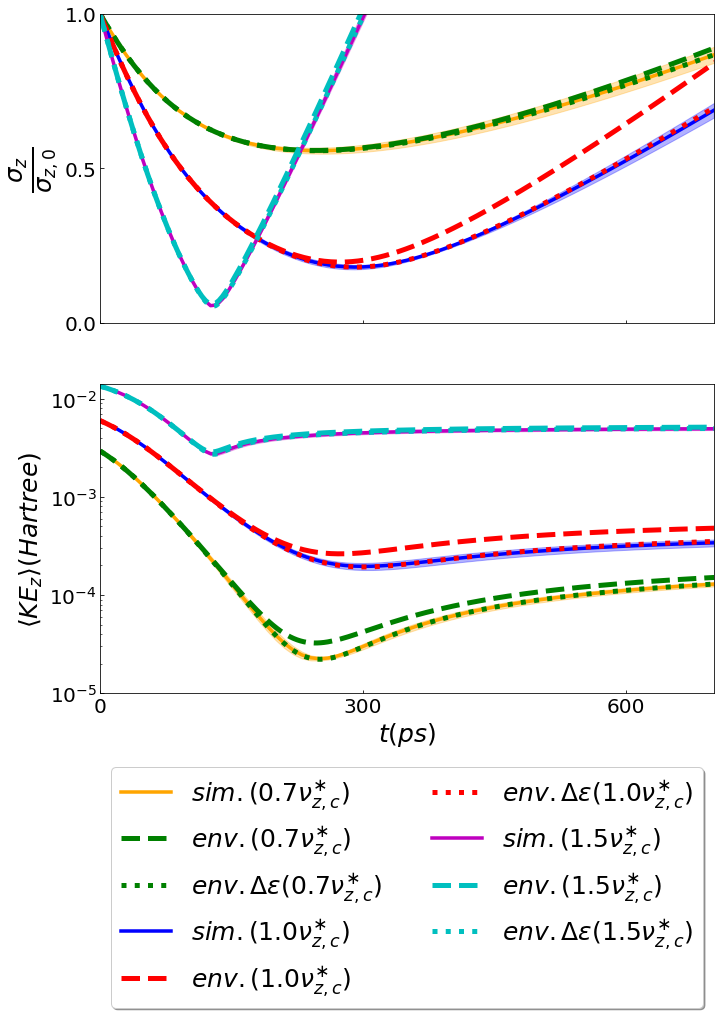}
	\end{center}
	\caption{\label{fig_MDSSA} Comparison of spatial width evolution and average kinetic energy in the longitudinal direction,
	$\frac{1}{N} KE_z = \frac{1}{2m} \left(\eta_z^2 +\frac{s_{z,p_z}^2}{s^2_z}\right) = \frac{1}{2m} s_{p_z}^2$, 
	of the bunch focused by three
	different initial chirps, $0.7 ~ \nu_{z,c}^\ast$, $1 ~ \nu_{z,c}^\ast$, and $1.5 ~ \nu_{z,c}^\ast$.  
	The line style of the plot indicates the simulations type: solid = mean of 90 N-particle simulations with the region shaded 
	within 1 standard deviation of the mean (sim), dashed = envelope
	equation with conserved emittance (env), and dotted = envelope equation with emittance provided from simulation 
	(env $\Delta \varepsilon$). 
	N-particle
	simulations were conducted by first thermalizing the bunch without chirp and with periodic boundary conditions before
	allowing the bunch to evolve with the appropriate chirp.
	The average initial phase-space statistics of the bunch post-thermalization were used to initialize the envelope equations.  
	Notice that the spatial envelope equations
	prediction is largely in agreement with the N-particle simulation except for the simulation at the critical chirp.
	For this simulation, the prediction deviates most significantly at the focal point.  
	Similar results can be seen with the kinetic energy evolution except notice deviation in the
	simulation in the bounce-back regime ($0.7 ~ \nu_{z,c}^\ast$).  
	These discrepancies are believed to be a result of the constant emittance assumption that is supported by re-examining 
	the envelope equations using the average emittance from the N-particle simulation at every time step (dotted lines).}
\end{figure}
As shown in Fig. \ref{fig_MDSSA}, the envelope equation predictions deviate from the $N$-particle 
simulations in three aspects: 1.) a slightly larger minimum width occurring at 2.) an earlier $t_f$ with 3.) a faster expansion after 
the focal point. These deviations are most significant at the critical chirp, where the bunch evolution is 
most sensitive to changes in emittance as we discussed in a previous section. We previously saw similar trends in the minimum width, 
the time to the minimum width, and the post focus expansion rate
as we increased the emittance in the envelope equations as can be seen in Fig. \ref{fig_Zt_SSA}. 
This suggests that the model used to predict the evolution of the width statistic 
likewise uses a larger longitudinal emittance than is seen in the $N$-particle simulation at crossover.  
As the envelope equations use the longitudinal emittance of the initial warm distribution that is 
used in the $N$-particle simulations, 
this suggests that the longitudinal emittance is in fact decreasing during the $N$-particle simulations. 
\begin{figure}
	\begin{center}
	\includegraphics[width=0.95\linewidth]{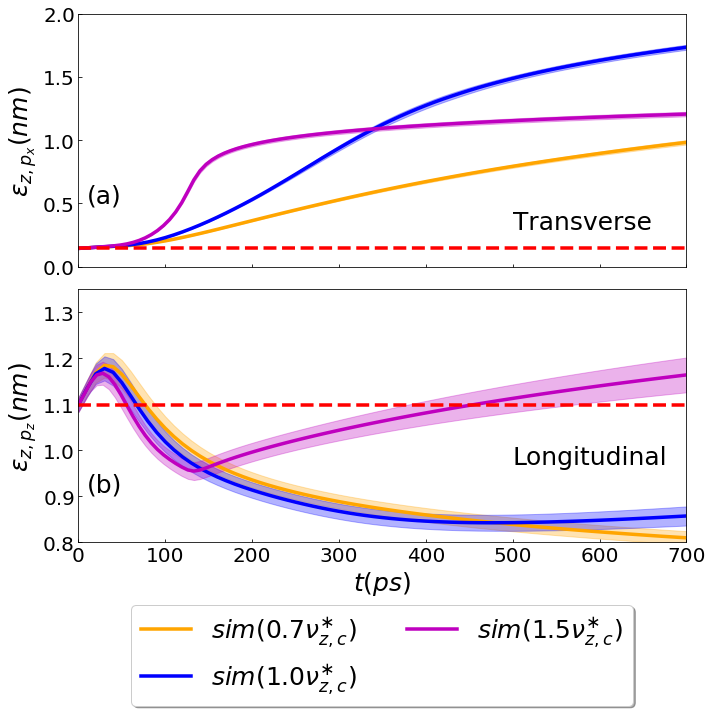}
	\end{center}
	\caption{\label{fig_emittanceFMM} (rms) Emittance evolution in (a) longitudinal $\varepsilon_{z,p_z}$ and (b) transverse $\varepsilon_{x,p_x}$ directions for the three different initial chirps, 
	$0.7 ~ \nu_{z,c}^\ast$ (orange), $1 ~ \nu_{z,c}^\ast$ (blue), 
	and $1.5 ~ \nu_{z,c}^\ast$ (magenta). All results were obtained from $N$-particle simulations.  Notice
	the initial bump in the longitudinal emittance within the first $100$ ps; this is driven by disorder induced heating that
	is not completely resolved by the protocol we used to thermalize the bunches.  Further notice that
	the longitudinal emittance decreases while transverse emittance increases after this point and before the focal point.  
	Also notice that the longitudinal emittance continues to
	decrease after the focal point for the chirps that are at or below the critical chirp; however, the longitudinal 
	emittance increases at and after the focal point in the crossover regime.  This occurs while the transverse emittance
	continues to increase, so the standard theoretical explanation of heat transfer between the dimensions does not
	describe this behavior and new theory is required to understand what is going on.  This is discussed further in the text.}
\end{figure}
Tracking the rms emittance of the $N$-particle simulations, as seen in panel (b) of Fig. \ref{fig_emittanceFMM},
confirms that the longitudinal emittance decreases prior to the focal point.

As discussed previously, the conservation of emittance in the envelope equations is a result of the term
$\frac{m^2 c^2}{2} \frac{d\varepsilon_{i,p_i}^2}{dt} = s_i^2 s_{p_i,F_i} - s_{i,p_i}s_{i,F_i}$ being $0$
in Eq (\ref{eq:eta}); conversely, the
non-conservation of emittance suggests that this term is non-zero.  Currently, there is no theory to predict
the value of these terms, but we can use the change in emittance seen in simulations in this term within the 
envelope equations to better capture the evolution.  Specifically, we replace
Eq. (\ref{eqn_SSA_eta}) by
\begin{equation} \label{eqn_SSAe_eta}
\frac{d \eta_i^2}{dt} =\frac{d}{dt}\left(\frac{\varepsilon^2_{i,p_i}}{\sigma^2_i} \right)  =  -\frac{2 \gamma_i \eta_i}{m\sigma^2_i} + \frac{1}{\sigma^2_i} \frac{d \varepsilon^2_{i,p_i}}{dt} 
\end{equation} 
in our envelope equations with $\frac{d \varepsilon^2_{i,p_i}}{dt} $ taken from the simulation results.
We note that this procedure was originally examined by Sacherer\cite{Sacherer:1971_envelope}.
The spatial width and longitudinal kinetic energy evolution
using these envelope equations with the simulation change in emittance squared term 
can be seen as the dotted lines in Fig.  \ref{fig_MDSSA}.
Excellent agreement between these modified envelope equations suggests
that varying emittance is the main factor causing the discrepancy between the longitudinal spatial variance and 
longitudinal kinetic energy evolution of the constant-emittance envelope equations and $N$-particle simulations. 
This suggests that if the physics of the covariance terms $s_{p_i,F_i}$ and $s_{i,F_i}$ can be understood and modeled,
that we should be able to obtain envelope equations that capture the expected behavior of electron bunches
to a high degree of accuracy. 

\section{\label{sec_FMMemittance} Discussion}
The emittance evolution in Fig. \ref{fig_emittanceFMM}, especially within the crossover regime
where the emittance in both the longitudinal and transverse directions increase simultaneously,
cannot be explained by the standard heat transfer mechanism employed in the literature.  We provide 
some insight into these dynamics here.
As can be seen in Fig \ref{fig_emittanceFMM}, the longitudinal emittance rapidly increases at the beginning of the simulation 
followed by a gradual decrease. For the simulations within the crossover regime, there is another rapid increase in the 
longitudinal emittance close to the focal point. In contrast, the transverse emittance has a rapid increase  
at the beginning of the simulation followed by a more gradual increase. Notice that there is again a 
rapid increase in the transverse emittance around the focal point. We emphasize that the rapid increase in the 
emittance of both directions is almost coincidental --- an observation not currently predicted from theory.

Within the literature, there seem to be two macroscopic ideas for the mechanisms involved in this process: 
\begin{enumerate}

\item \textit{Emittance transfer between degrees of freedom (Heat transfer):}    As
emittance can be thought as proportional to the square root of the heat times the spatial extent, this heat
transfer results in emittance transfer between the degrees of freedom.

\item \textit{Disorder-induced heating (DIH):} DIH in the plasma community describes the heating process during the transition 
from a disordered state to one which is structured by Coulomb 
forces. 

\end{enumerate}
We point out that these two ideas have previously been described in the literature.  Specifically,
Reiser's standard book in accelerator physics describes the heat transfer\cite{Reiser:1994_book}, and 
DIH has phenomenologically been described by Gericke et. al and Maxson et. al\cite{Gericke:2003_dih, Maxson:2013_DIH}.  
Further, Struckmeier discussed these two ideas in his work on modeling envelope
equations with additional Fokker-Planck style random terms\cite{Struckmeier:1994_fokker_planck,Struckmeier:1996_entropy,Struckmeier:2000_stochastic} with slightly different language and a more mathematical presentation; 
however, it is not clear to us if these effects are truly stochastic
in the manner that should be modeled by Fokker-Planck statistics.  We also point out that these ideas are mechanisms
in the language of thermodynamics but do not describe mechanisms in the statistical physics sense as
their definition does not lead to any inherent time scale.  

Nonetheless, using these two ideas and our own notation, we can phenomenologically explain the emittance evolution seen in 
Fig. \ref{fig_MDSSA}.  First we define the linear heat along $i$ by $\frac{1}{2m} \eta_i^2$,
which can be viewed as the kinetic energy contained in the random fluctuations (from linearity), and
we use this measure as a proxy for the heat of the distribution.  Recall 
that the emittance can be written as $\varepsilon_{i,p_i} = \frac{1}{mc} s_i \eta_i$, which can be thought as being proportional to
the product of the spatial width and the local momentum width or equivalently the square root of the linear heat. 
At the beginning of the simulation after the confinement is suddenly removed, potential energy is released through DIH into $\eta_T$
and $\eta_z$ while $s_T$ and $s_z$ do not change much.  This results in a sudden increase in both $\varepsilon_{x,p_x}$
and $\varepsilon_{z,p_z}$ that continues even as $s_T$ and $s_z$ begin to significantly change.  

Next, the longitudinal emittance begins to decrease as the transverse emittance continues to increase.  
To understand this phenomenologically, we first consider how the linear heat would change under emittance conserving conditions.
As $\eta_z = \frac{\varepsilon_{z,p_z}}{s_z}$, 
a decrease in $s_z$ would result in an increase in $\eta_z$
 under the assumption of 2D conserved emittance.  Likewise,
$\eta_x = \frac{\varepsilon_{x,p_x}}{s_x}$, and an increase in $s_x$ would result in a decrease in $\eta_x$.  
In other words if the 
2D emittance were conserved, we'd expect the linear heat in the longitudinal direction,
i.e. $\frac{1}{2m}\eta_z^2$, to increase, and we'd expect the linear heat in the transverse 
direction , i.e. $\frac{1}{2m}\eta_x^2$, to decrease.  
Notice that a strict definition of temperature is not appropriate for our highly non-equilibrium situation, but that this definition of 
linear heat is still appropriate.  
As the initial thermalization leads to these two heats being roughly the same, the
difference in the heat that develops as the longitudinal dimension contract and the transverse dimension expands
would lead to the development of a thermal gradient between the longitudinal and transverse
dimensions. 

\begin{figure}
	\begin{center}
	\includegraphics[width=0.95\linewidth]{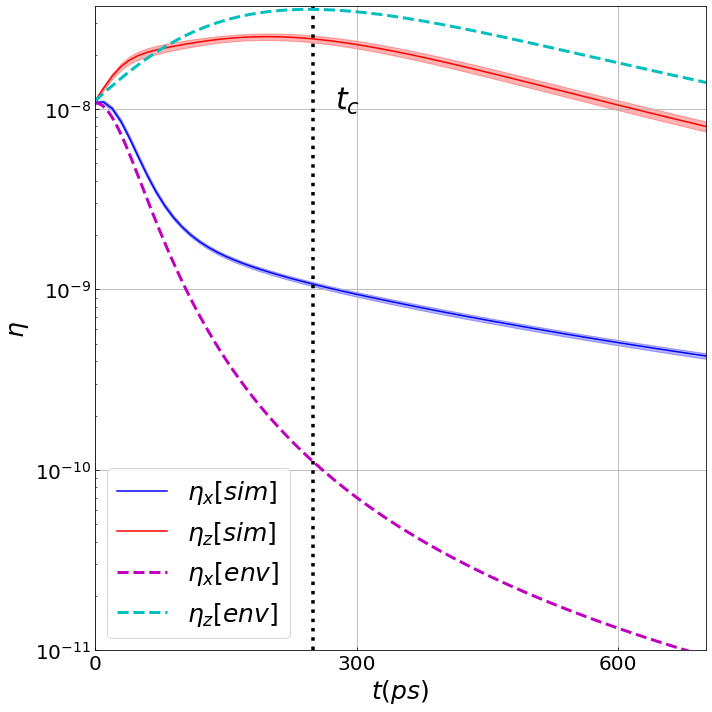}
	\end{center}
	\caption{\label{fig:eta ev} Comparison of the evolution of the parameter $\eta$ between the 
	envelope equations (dashed lines) 
	and the $N$-particle simulations (solid lines).  The line and shaded area around the $N$-particle simulation lines
	represent the mean (solid line) $\pm$ the standard deviation of 90 simulations.  The vertical dotted line indicates the
	focal point.
	Notice that the emittance conserving model
	over-predicts $\eta_z$ and under-predicts $\eta_x$.  This is partially due to the heat being transferred between the dimensional
	modes -- a mechanism that is not captured by the emittance conserving envelope equations.}
\end{figure}

Now in the non-emittance conserving condition, heat can be transferred between the dimensions.  So as the bunch focusses,
we'd expect a heat transfer from the hotter longitudinal to the cooler transverse direction --- that is, the evolution
of simulated $\eta_z$ would be expected to be smaller than the $\eta_z$ we'd obtain from a the emittance preserving 
envelope equations; conversely, we'd expect the simulated $\eta_x$ to be larger than the theory $\eta_x$.  This is precisely
what is seen in Fig. \ref{fig:eta ev}.
This in turn results
in the longitudinal emittance decreasing while the transverse emittance increases, which is precisely what happens in the
bounce-back regime.  

However, once the bunch is in the crossover regime, there is actually an increase in the longitudinal emittance 
in Fig \ref{fig_emittanceFMM}.  This phenomenon could be explained from this perspective by postulating that a second
period of DIH occurs near the focal point when the bunch is within
the crossover regime.  That is, if the particles crossover, they are forced into a highly non-equilibrium state that rapidly relaxes releasing heat into the bunch thus increasing the emittance.

\section{Conclusions}
In this work, we have examined the longitudinal crossover of electron bunches with uniform ellipsoidal profiles focused by a linear chirp as is typical of the propagation of a probing electron bunch in an ultrafast electron diffraction/microscope system. 
We employed several analytic techniques to model the space charge dynamics of the bunch, 
the first of which is an extension of Grech et. al's mean-field theory which utilizes ordinary differential equations for the ellipsoid transverse and longitudinal sizes to describe the bunch evolution.  Analysis of this mean-field model leads to the identification of a longitudinal critical chirp. This critical chirp separates two regimes for particle trajectories in this model: bounce-back, where the particles reverse their direction at the bunch waist, and crossover, where the bunch experiences a singularity with a width of zero. 
We showed that time can be scaled by the initial plasma frequency, and by defining a dimensionless critical chirp the zero-emittance model behavior depends only on the initial aspect ratio. The evolution of bunches with the same initial geometry then differ only by the time scale determined by the bunch's plasma frequency.

We examined the problem through the statistical formulation of envelope equations by building on Sacherer's statistical analysis of the cylindrical KV-envelope equations 
that are well known in the accelerator physics community.  In other work, we recently presented a statistical perspective on the envelope
equations that we are calling the sample perspective\cite{Zerbe:2020_1d_emittance}. 
We showed that the statistical envelope equations for three dimensional systems are identical to, 
up to a constant we determine in Appendix \ref{sec_AG_appx}, the  Analytical Gaussian formalism that is well known in the 
ultrafast electron microscopy community; the envelope equations from the sample perspective are more general
and have a more straightforward derivation than Michalik and Sipe's integral approach.  
It should be noted that the majority of analysis of envelope equations in the accelerator physics community is in 
cylindrical coordinates due to the predominance of accelerators with continuous beams or which have bunches with very 
large prolate aspect ratios.  
In contrast in the UEM/UED field, the bunch near the source and at the longitudinal focal point is a highly oblate ellipsoid, 
or pancake, and a fully three dimensional description is required.   
Further, we showed that the mean-field theory utililzed in the Coulomb explosion literature, and extended here,  is simply 
the zero-emittance limit of these envelope equations.  Moreover the zero emittance or modified Coulomb explosion model 
yields a more detail analytic treatment, and it clearly reveals the critical chirp phenomenon 
separating the bounce back and crossover regimes at the bunch focal point described above. 

As currently formulated, the envelope equations rely on the assumption of a linear force as applies to uniform ellipsoidal bunches, 
and they also rely on the assumption of a linear relation between average momentum and average position in each of the 
three directions.  Following the convention in the literature, we call this a linear chirp assumption.  The envelope 
equations developed by Sacherer describe deviations from the linear chirp which may be either global or statistical non-linearities. 
The power of this method is that it describes arbitrary distributions, provided the form of the distribution does not change.  
It therefore encompasses the uniform ellipsoid theory of the KV equations, the Analytic Gaussian model and any 
other distribution with well defined second moments. 
An important result is that the emittance is conserved in the envelope equations if the linear force, 
linear chirp and conserved distribution assumptions are utilized.   Nevertheless, Sacherer 
realized\cite{Sacherer:1971_envelope} that a changing emittance is sometimes observed and that this could 
also be included in the envelope equations.   
The N-particle simulations carried out here demonstrate clearly that emittance is not conserved near the electron bunch 
longitudinal focal point and that if the computationally determined emittance dynamics is included in the envelope equations 
then the envelope equations provide an accurate description of the longitudinal and transverse sizes of the bunch through 
the focal point as should be expected.    This provides an important formulation that should be very useful in understanding 
and designing the longitudinal focusing systems for UEM and UED microscopes. Furthermore, 
we recently showed\cite{Zerbe:2020_1d_emittance} 
that the statistical kinematics, also originally derived by Sacherer to obtain his envelope equations, are completely general and exact
and therefore can be used to describe the bunch evolution with no assumptions about the chirp, initial profile, 
or distribution evolution. We expect that better understanding of these equations and of emmitance growth and transfer can be used to obtain better models of the dynamics of high intensity charged particle bunches and their applications to both 
UEM systems and to other accelerator physics applications. 

We provided a qualitative description of the emittance growth and transfer observed in Fig. 6 by elucidating three mechanisms: 
(i)  Disorder induced heating, or relaxation from a highly non-equilibrium state, which converts potential energy into kinetic 
energy and naturally leads to a rapid growth in emittance.   This effect occurs at the start of our simulations and near the focal point.  
(ii)  Transfer of``heat", which is equal to kinetic energy fluctuations,  from the hotter direction to the colder direction.  
(iii) The fact that the kinetic energy fluctuations increase along a direction that is compressed and decrease along a 
direction that is expanding.   This last mechanism allows for one direction to become hotter than the other colder, 
e.g. when the longitudinal direction is being compressed it becomes hotter while the transverse direction cools down as it expands.  
The system then tries to equilibrate by transferring heat from the hot direction to the cold direction.   
We note that the longitudinal emittance, the longitudinal momentum fluctuations and the longitudinal bunch size are related by 
$\epsilon_{z, p_z} = s_z \eta_z$ and that the longitudinal non-equilibrium heat is given by $\eta_z^2/2m$.   
Therefore dynamics of the heat, or momentum fluctuations, and the dynamics of the emittance may be different due to the $s_z$ factor as is evident when comparing Figs. \ref{fig_emittanceFMM} and \ref{fig:eta ev}.

In conclusion, by combining analytic methods from several fields of research with state of the art N-particle simulations we are able 
to present a accurate theory to describe the evolution of electron bunch aspect ratio through the focal point.  The 
N-particle simulations provide new insights into the mechanisms for emittance growth and transfer that challenge us to 
develop new theories to capture these mechanisms quantitatively.  This challenge and the extension of the theories 
developed here to the relativistic regime define productive directions for future work.  

\begin{acknowledgments}
This work was supported by NSF Grant numbers RC1803719 and RC108666.  We thank Steve Lund, Reginald Ronningen,
Chong-Yu Ruan, Carl Schmidt, and David Tomanek for their advice.
\end{acknowledgments}

\appendix
\begin{center}
	\textbf{Appendix}
\end{center}

\section{Relation between AG model and envelope equations\label{sec_AG_appx}}
Now that is is apparent that the assumptions of the AG model lead to a linear force on the ensemble particles, we show that
the evolution of the AG model is equivalent to the envelope equations assuming a uniform distribution up to a constant.  Specifically,
we show that the the force portion of Eq. (\ref{eq:gamma}) obtained by Michalik and Sipe by integration techniques is the same up to
a constant to the analogous term obtained by using the mean-field force within a uniform ellipsoid.  Knowing that
$s_{i,k_i i} = k_i s_i^2$ we infer slopes of the force for Michalik and Sipe to
\begin{align}
  k_i^{MS} = \frac{1}{4\pi\epsilon_0} \frac{N e^2}{6 \sqrt{\pi} s_i^3} L_i\left(\frac{s_z}{s_T}\right)
\end{align}
where 
\begin{subequations}\label{eq:gauss int}
\begin{align}
  L_z(a) &= \frac{3a^2}{a^2-1}\left(a L(a) - 1\right)\\
  L_T(a) &= \frac{3}{2}\left(L(a) + \frac{a^2 L(a) - a}{1-a^2}\right)\\
  L(a) &= \begin{cases}
                 \frac{arcsin \sqrt{1-a^2}}{\sqrt{1 - a^2}},& 0 \le a \le 1\\
                 \frac{\ln(a + \sqrt{1 - a^2})}{\sqrt{a^2-1}},& 1\le a
               \end{cases}
\end{align}
\end{subequations}
On the other hand, the slopes of the force for a uniform ellipsoid with $s_x = s_y$ can be written as
\begin{subequations}
\begin{align}
  k_x^{unif} &= \frac{1}{4 \pi \epsilon_0} \frac{ 3 N e^2 }{10 \sqrt{5} s_x^3} \beta\left(1,\frac{s_z}{s_x}\right)\\
  k_z^{unif} &= \frac{1}{4 \pi \epsilon_0} \frac{ 3 N e^2 }{10 \sqrt{5} s_x^3} \beta\left(\frac{s_x}{s_z},\frac{s_x}{s_z}\right)\\
\end{align}
\end{subequations}
where 
\begin{align}
  \beta(a,b) &=  \int_0^{\infty} \frac{1}{(1+u)^{3/2}\sqrt{a^2+u}\sqrt{b^2+u}} du \label{eq:unif int}
\end{align}
A derivation of this is presented in our recent work\cite{Zerbe:2020_1d_emittance}.
The comparison between these two models comes down to how Eq. (\ref{eq:gauss int}) and Eq. (\ref{eq:unif int}) compare.
Specifically, letting $a = \frac{s_z}{s_x}$ as it is in Eq. (\ref{eq:gauss int}), we need to evaluate $\beta(1,a)$
and $\beta(\frac{1}{a},\frac{1}{a})$ and then compare the two slopes.

We start the evaluation of the transverse relevant integral:
\begin{align}
  \beta(1,a) &= \int_0^{\infty} \frac{1}{(1+u)^{2}\sqrt{a^2+u}} du\nonumber\\
                   &= \frac{cos^{-1}(a) - a\sqrt{1-a^2}}{(1-a^2)^{3/2}}\nonumber\\
                   &= \begin{cases}
                          \frac{ sin^{-1} (\sqrt{1 - a^2})- a\sqrt{1-a^2}}{(1-a^2)^{3/2}},& 0\le a \le 1\\
                          \frac{ i \ln\left(a + \sqrt{a^2 -1}\right)- a\sqrt{1-a^2}}{(1-a^2)^{3/2}},&  a \ge 1
                        \end{cases}\nonumber\\
                   &= \begin{cases}
                          \frac{ sin^{-1} (\sqrt{1 - a^2})}{\sqrt{1-a^2}(1-a^2)} - \frac{a}{1-a^2},& 0\le a \le 1\\
                          \frac{ \ln\left(a + \sqrt{a^2 -1}\right)}{\sqrt{a^2-1}(1-a^2)} - \frac{a}{1-a^2},&  a \ge 1
                        \end{cases}\nonumber\\
                    &= \frac{L(a) - a}{1-a^2}\nonumber\\
                    &= L(a) + \frac{a^2 L(a) - a}{1-a^2}\nonumber\\
                    &= \frac{2}{3} L_T(a)
\end{align}
Thus, our uniform integrals in the transverse direction differs by a factor of $\frac{2}{3}$ from the AG model's Gaussian integrals
in the transverse direction.

Next we evaluate the longitudinally relevant integral:  
\begin{align}
  \beta\left(\frac{1}{a},\frac{1}{a}\right) &= \int_0^{\infty} \frac{1}{(1+u)^{3/2}(\frac{1}{a^2}+u)}\nonumber\\
                  &= \frac{2}{\frac{1}{a^2} -1} - \frac{2 sec^{-1} \left(\frac{1}{a}\right)}{(\frac{1}{a^2} - 1)^{3/2}}\nonumber\\
                  &= \frac{2a^2}{\left(1 - a^2\right)} \left(1 - a \frac{cos^{-1} (a)}{\sqrt{1 - a^2}}\right)\nonumber\\
                  &= \begin{cases}
                           \frac{2a^2}{\left(a^2 - 1\right)} \left(a \frac{sin^{-1} (\sqrt{1 - a^2})}{\sqrt{1 - a^2}} - 1\right),& 0\le a \le 1\\
                           \frac{2a^2}{\left(a^2 - 1\right)} \left(a \frac{i \ln\left(a + \sqrt{a^2 -1}\right)}{\sqrt{1 - a^2}}- 1\right),&  a \ge 1
                        \end{cases}\nonumber\\
                  &= \begin{cases}
                           \frac{2a^2}{\left(a^2 - 1\right)} \left(a \frac{sin^{-1} (\sqrt{1 - a^2})}{\sqrt{1 - a^2}} - 1\right),& 0 \le a \le 1\\
                           \frac{2a^2}{\left(a^2 - 1\right)} \left(a \frac{\ln\left(a + \sqrt{a^2 -1}\right)}{\sqrt{a^2-1}}- 1\right),& a \ge 1
                        \end{cases}\nonumber\\ 
                    &= \frac{2}{3} L_z\left(a\right)     
\end{align}
Again we see that the same factor is present between the integrals in the longitudinal direction, which is reassuring.

Putting this factor of $\frac{2}{3}$ into the comparison of the slopes, we see that the slopes are related by
\begin{align}
  \frac{k_i^{MS}}{k_i^{unif}} &= \frac{5\sqrt{5}}{6\sqrt{\pi}} \approx 1.05
\end{align}
So we see that these two models differ in their forces only by roughly 5\%.  
This small difference in the 
linear force should result in either model adequately capturing the dynamics of either uniform or Gaussian evolution if the
changes in emittance, that should be more prevalent in the Gaussian model, are ignored.  
Furthermore, even this difference may be absorbed
by the assumed number of particles when fitting parameters.  
That is, in both modes $N$ needs to be set.  As the only term that depends
on $N$ is the force, setting the $N$ in the uniform envelope equations 5\% larger than the $N$ in the AG model will result in the
same exact solution. So in essence the AG model is identical to the uniform envelope equation but with a slightly adjusted number of 
particles.

\bibliographystyle{apsrev4-1}
\bibliography{xukuns_paper.bib}

\end{document}